\documentclass[]{elsart}
\usepackage{graphicx,natbib,amssymb}

\def\rc{\mbox{$R_{\rm core}$}}
\def\ms{\mbox{$M_\odot$}}
\def\ds{\mbox{$d_\odot$}}

\def\ks{\mbox{$K_s$}}

\def\arcmin{\mbox{$'$}}
\def\arcsec{\mbox{$''$}}

\begin{document}

\begin{frontmatter}

\title{The embedded cluster DBSB\,48 in the nebula Hoffleit\,18: comparison with 
Trumpler\,14\thanksref{label1}}

\thanks[label1]{Observations collected at the New Technology Telescope (NTT), European Southern 
Observatory (ESO), La Silla, Chile, proposal 071.D-0506(A).}

\author[lbl1]{S. Ortolani}
\author[lbl2]{C. Bonatto\corauthref{cor1}}
\ead{charles@if.ufrgs.br}
\ead[url]{www.if.ufrgs.br/$\sim$charles}
\corauth[cor1]{Tel. 55 51 3308-6432; FAX 55 51 3308-7286}
\author[lbl2]{E. Bica}
\author[lbl1]{Y. Momany}
\author[lbl3]{B. Barbuy}

\address[lbl1]{Universit\`a di Padova, Dipartimento di Astronomia, Vicolo dell'Osservatorio 2,
I-35122 Padova, Italy}

\address[lbl2]{Departamento de Astronomia, Universidade Federal do Rio Grande do Sul, \\
Av. Bento Gon\c{c}alves 9500, Porto Alegre 91501-970, RS, Brazil}

\address[lbl3]{Universidade de S\~ao Paulo, Rua do Mat\~ao 1226, 05508-900, S\~ao Paulo, Brazil}

\begin{abstract}
We derive fundamental parameters of the embedded cluster DBSB\,48 in the southern nebula Hoffleit\,18 
and the very young open cluster Trumpler\,14, by means of deep JH\ks\ infrared photometry. We build 
colour-magnitude and colour-colour diagrams to derive reddening and age, based on main sequence 
and pre-main sequence distributions. Radial stellar density profiles are used to study cluster 
structure and guide photometric diagram extractions. Field-star decontamination is applied to 
uncover the intrinsic cluster sequences in the diagrams. Ages are inferred from K-excess fractions.
A prominent pre-main-sequence population is present in DBSB\,48, and the K-excess fraction 
$f_K=55\pm6\%$ gives an age of $1.1\pm0.5$\,Myr. A mean reddening of $A_{K_s}=0.9\pm0.03$ was found, 
corresponding to $A_V=8.2\pm0.3$. The cluster CMD is consistent with the far kinematic distance 
of 5\,kpc for Hoffleit\,18. For Trumpler\,14 we derived similar parameters as in previous studies 
in the optical, in particular an age of $1.7\pm0.7$\,Myr. The fraction of stars with infrared excess 
in Trumpler\,14 is $f_K=28\pm4\%$. Despite the young ages, both clusters are described by a King 
profile with core radii $\rc=0.46\pm0.05$\,pc and $\rc=0.35\pm0.04$\,pc, respectively for DBSB\,48 
and Trumpler\,14. Such cores are smaller than those of typical open clusters. Small cores are probably 
related to the cluster formation and/or parent molecular cloud fragmentation. In DBSB\,48, the magnitude 
extent of the upper main sequence is $\Delta\,\ks\approx2$\,mag, while in Trumpler\,14 it is
$\Delta\,\ks\approx5$\,mag, consistent with the estimated ages.
\end{abstract}


\end{frontmatter}

\section{Introduction} 
\label{Intro}

Infrared clusters represent a new class of objects, virtually undetectable before the 90's (e.g. 
\citealt{Dehar97}; \citealt{Hodapp94}). Until recently, the number of known infrared 
clusters and stellar groups amounted to 276, as shown in the compilation by \citet{BiDuBa03}. 
A systematic survey by \citet{DuBiSoBa03} and \citet{BiDuSoBa03} in directions of nebulae using
2MASS\footnote{The Two Micron All Sky Survey, All Sky data release \citep{Skru97}, available at 
{\em http://www.ipac.caltech.edu/2mass/releases/allsky/}} revealed 346 new infrared embedded clusters 
and candidates. Thus, the systematic study of embedded clusters is fundamental to understand their 
nature, to probe the physical conditions associated to the early stages of star clusters, and to 
derive their fundamental parameters. Just to mention a few efforts in that direction, Serpens 
\citep{OT02}, NGC\,1333 \citep{Warin96}, NGC\,3576 \citep{Persi94}, AFGL\,5142 \citep{CNSS93}, and 
S\,106 \citep{HodRay91}. 


Embedded clusters may provide the clues  to better understand the formation and evolution processes 
of star clusters and their interaction with the parent molecular cloud. Colour-colour diagrams (2-CDs) 
allow identification of Pre-Main-Sequence (PMS) stars and can be used to distinguish them from Main 
Sequence (MS) and field stars (\citealt{CNSS93}; \citealt{LL03}). Together with colour-magnitude 
diagrams (CMDs) they can be used to derive reddening values, reddening distribution, distance from
the Sun and age. K-band infrared excesses originate mostly in dust envelopes and/or discs of PMS stars, 
and indicate their evolutionary stage up to $\sim10$\,Myr (\citealt{Greaves05}; \citealt{MB05}), or 
more \citep{BoBiOrBa06}. K, and particularly L, excesses are sensitive to the presence of protoplanetary 
discs (\citealt{HLL01a}; \citealt{OJL04}). Thus, these indicators allow dating very young star clusters 
(e.g. \citealt{LAL96}; \citealt{SoBi03}). In this context, it is important to increase the number of 
embedded clusters studied in detail in order to establish star-formation age spreads and constrain survival
time-scales of dust envelopes and circumstellar discs.

Deriving locations of embedded clusters in the Galaxy by means of photometric and spectroscopic 
methods is useful, since these estimators might provide distances to be compared with the available 
kinematic ones of the nebulae (\citealt{GG76}; \citealt{BFS82}). Embedded clusters are 
typically observed up to $\sim4$\,kpc from the Sun \citep{LL03}. In turn, these studies can be used to
build the Galactic structure around the Sun, in particular to trace spiral arms. Improvement in distance 
determinations for clusters in all disc directions would contribute to the derivation of more reliable 
parameters of the rotation curve of the Galaxy. The rotation curve and spiral structure has been
studied by \citet{Rus03}.

In this paper fundamental cluster parameters are derived using infrared photometry 
of the embedded cluster DBSB\,48 and the young open cluster Trumpler\,14.

Hoffleit\,18 is a southern nebula discovered by \citet{Hof53}. The cluster embedded in this 
nebula is Dutra, Bica, Soares, Barbuy\,48 - DBSB\,48 - \citep{DuBiSoBa03}, located at J2000 $\alpha=10^h31^m29^s$, 
$\delta=-58^\circ02\arcmin01\arcsec$ ($\ell=285.26^{\circ}$, $b=-0.05^{\circ}$). The 
estimated diameter of the cluster is 1.5\arcmin.

The nebula Hoffleit\,18 was also identified as a radio H\,II region designated G285.3+0.0 or 
G285.253-0.05. The derived radio velocity is $-2\rm\,km\,s^{-1}$, implying in this direction a near 
distance of 0.3 kpc and a far distance of 5.0 kpc. Given that the object is faint, the far distance 
should be the correct one, placing it in the Sagittarius-Carina arm \citep{CaHa87}. In the present 
paper we check the consistency of the CMD loci with the kinematic distance estimate. 

Trumpler\,14 is usually classified as a young open cluster (\citealt{VBF96}, and references 
therein). Besides, it is embedded in the nebula NGC\,3372 ($\eta\,\rm Car\ Nebula$) and its optical
populous nature and environment properties make it an ideal object to be compared to a bona-fide 
embedded cluster such as DBSB\,48. Trumpler\,14 contains about 13 O stars, and is a relatively 
massive cluster, with a mass estimated to be around $2000\,\ms$ \citep{VBF96}. 

In Sect.~\ref{ObsDR} the observations are described. In Sect.~\ref{DBSB48} the structure, CMD and 
2-CDs of DBSB\,48 are discussed, field-star decontamination is applied to the CMDs, and fundamental
parameters are derived. Trumpler\,14 is dealt with similarly in Sect.~\ref{Tr14}. Discussions and 
concluding remarks are in Sect.~\ref{Conclu}.

\section{Observations and data reductions} 
\label{ObsDR}
 
The observations were carried out in 2003, May 17-18, using the  SOFI Camera \citep{sofi}  at the 
3.55m New Technology Telescope (NTT) at ESO, La Silla. The detector was a HAWAII Rockwell HgCdTe with 
$1024\times1024$ pixels, with a pixel size of $18.5\,\mu$m. We used the large field mode with a projected 
field of view of $4.95\arcmin\times4.95\arcmin$. The projected size of the pixel on the sky is 0.292\arcsec.
DBSB\,48 was observed on the night of May 17. Total exposure times of 12 min in J and H, and 15 min in \ks\ 
have been obtained in the cluster and background regions.  The average seeing  was 
$\simeq0.9\arcsec$ in $J$ and $H$ bands, and $\simeq0.7\arcsec$ in the \ks\ band. Short exposures 
of 1.2 sec for a total 
of 60 sec were also taken in the three filters. Further details on observations, reductions and 
calibrations with this instrumentation can be found in \citet{DuOrBiBaZM03}. The night was photometric 
and calibrations were obtained with 6 standard stars from \citet{Persson98}, and they were 
checked with 2MASS photometry. For the last filter the calibration was made in \ks. For the 
photometric reductions the DAOPHOT code \citep{Stet87} was used in IRAF environment. 

\begin{figure}
\resizebox{\hsize}{!}{\includegraphics{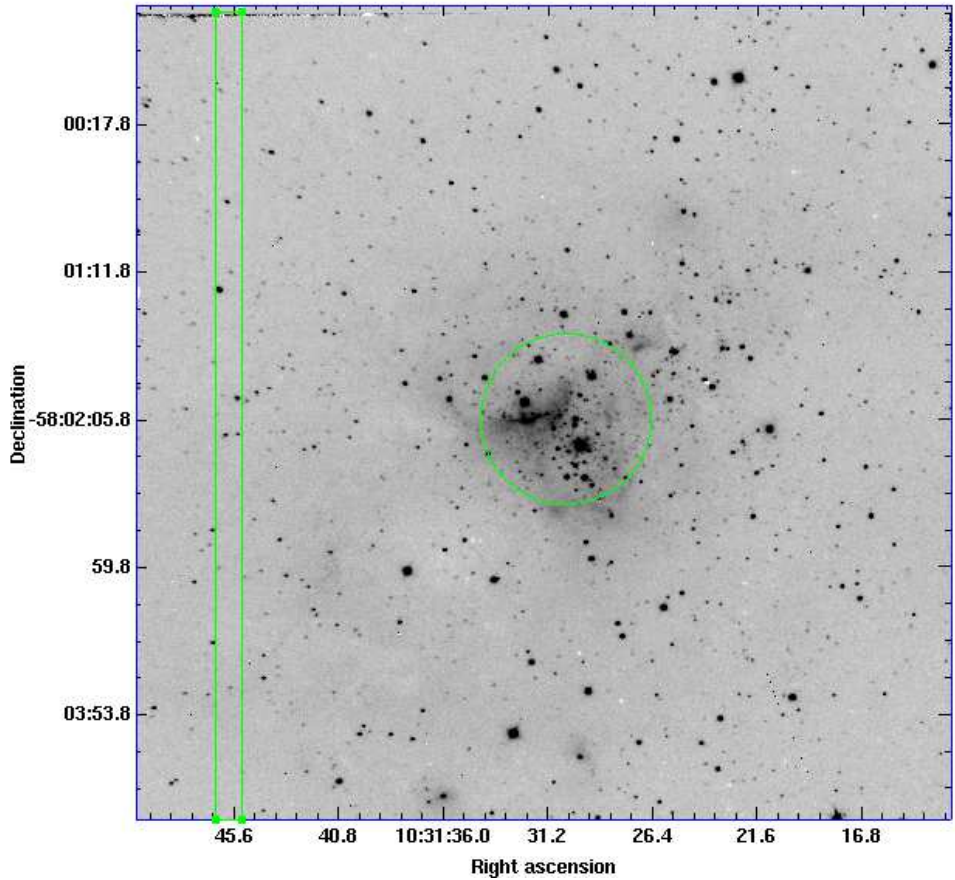}}
\caption{DBSB\,48: \ks\ image showing an extraction of $4.9\arcmin\times4.9\arcmin$. North is up 
and east left. A circle indicating the region encompassed by a radius $r<26\arcsec$\ is shown. The
comparison field is shown by a rectangle.}
\label{fig1}
\end{figure}

\begin{figure}
\resizebox{\hsize}{!}{\includegraphics{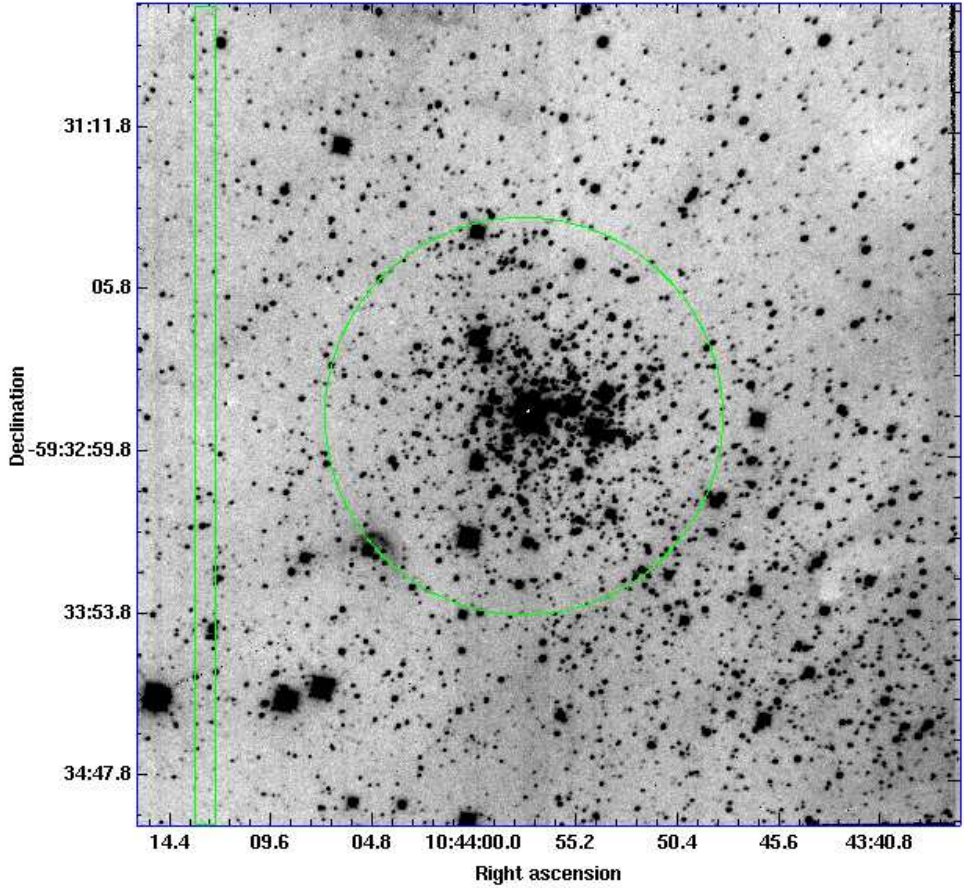}}
\caption{Same as Fig.~\ref{fig1} for the \ks\ image of Trumpler\,14.}
\label{fig2}
\end{figure}

Trumpler\,14 was observed on May 18 for comparison purposes. The total exposure time on object 
and sky was 500 sec per  band, plus multiple short exposures for a total of 60\,sec per band. 
 The average  seeing  was  $\simeq1.2\arcsec$  in  $J$  and $H$  bands   and
$\simeq0.9\arcsec$ in the \ks\ band. Because the night was not photometric, the calibration 
was obtained via 2MASS photometry. 

 Fig.~\ref{fig1} shows a \ks\ image of the cluster DBSB\,48, where traces of the nebula 
Hoffleit\,18 are seen, probably dominated by its reflection component in the infrared. In 
Fig.~\ref{fig2} the \ks\ image of Trumpler\,14 shows traces of emission from the $\eta$ Carina 
Nebula.
       
Photometric uncertainties from DAOPHOT, calculated from Poisson statistics, are provided 
in Table~1. The total photometric errors are larger because they also include uncertainties  
due to aperture correction, blends, non-linear pixels and flat field residuals.
 Below we include  a brief description on  the reduction and calibration of SOFI data. 
A detailed  presentation can be found in \citet{DuOrBiBaZM03} and \citet{Momany03}.

The pre-reduction of SOFI data (dark-frame subtraction, sky subtraction and flat fielding)
was performed following the steps given in the SOFI manual (\citealt{Lidman00}). In the  
process of flat-fielding the illumination correction frames and the bad pixel maps, both 
available from the ESO web-pages, were used. SOFI is known to be linear up to about 5000 
A.D.U. at a level of $0.1\%$, showing a $1.0\%$ non-linearity at 13000-14000 A.D.U. In this
sense, we note that our observing strategy (taking both shallow and deep exposures) of  
both clusters guaranteed  that the raw science images (i.e. including the sky level) should 
not require linearity correction.

Compared with globular clusters, DBSB\,48 and Hoffleit\,18 are relatively non-crowded.  
Nevertheless, we made use of the DAOPHOT II and ALLFRAME (\citealt{Stet87}, \citealt{Stet94})  
photometric reduction packages (designed for crowded fields). This ensured the derivation of 
PSF magnitudes for faint stars, even in the central regions of DBSB\,48 and Hoffleit\,18.
It is important  to note   that  our ALLFRAME  reduction (which  makes
simultaneous use of the geometric and photometric information from all
images)  thus extends the range of  magnitude  and crowding conditions
for which useful photometry  is obtainable.
Once the FIND and PHOT tasks  were performed, we searched for isolated
stars  to build the   PSF for each  single image.   The final PSF were
generated with  a PENNY function  that had  a quadratic  dependence on
position  in the frame.   The final instrumental photometric catalogue
was constructed (using the DAOMASTER  package) by scaling the  shallow
and deep magnitudes  to a common  photometric reference, and averaging
the magnitudes for the stars in common.
This catalogue contains PSF  magnitudes,  and these were converted  into
aperture magnitudes  assuming  that $m_{\rm ap}=m_{\rm  PSF}-constant$
(\citealt{Stet87}), where the constant  is the aperture correction that we
derived separately using bright isolated stars in the field.

The  observation night of  DBSB\,48 was photometric  and the photometric
calibration was  obtained using 6  standard stars from  \citet{Persson98}.  
An   independent  check of our  photometric   calibration was
obtained from  a comparison with  2MASS  photometry.  This  showed the
presence of  no systematic effects,  and yielded differences among the
brightest stars   within   $\sim0.035$ for  the   $JHK_{s}$ magnitudes
scales.

To  assess the photometric  errors  and the completeness we  performed
artificial star experiments  only for the DBSB\,48  data set.  Simulated
stars were  added  to the sum  image  (used  in ALLFRAME) by  randomly
distributing them  around   the nodes of  a  grid   of triangles  -  a
procedure well suited to prevent self-crowding (see also \citealt{Momany02}).   
Stars   were simulated at    a fixed  intermediate  colour  of
($J-\ks$)$=1.5$, and varying \ks\ magnitudes.   The frames (with    the
artificial stars)  were then reduced and  calibrated  as done  for the
original images.   The artificial stars  experiments indicate  that at
$\ks\sim18.0$ we reach a photometric completeness level of $90\%$. 

For cluster analysis we applied a soft error filter to avoid the discrepant (mostly spurious) faint 
detections. Stars with errors larger than $\epsilon(J)=0.2$, $\epsilon(H)=0.25$ and $\epsilon(\ks)=0.3$ 
were discarded.


\begin{table}
\caption[1]{Photometric uncertainties}
\begin{flushleft}
\begin{tabular}{lllllll}
\noalign{\smallskip}
\hline
\noalign{\smallskip}
J range & $\epsilon(J)$ & H range  & $\epsilon(H)$ & \ks\ range & $\epsilon(\ks)$  \\
\noalign{\smallskip}
\hline
\noalign{\vskip 0.2cm}
J$<$16 & 0.014 & H$<$16 & 0.03 & $\ks<$16 & 0.02 \\
16$<$J$<$18 & 0.03 & 16$<$H$<$18 & 0.06 & 16$<\ks<$18 & 0.09 \\
18$<$J$<$19 & 0.09 & 18$<$H$<$19 & 0.17 & 18$<\ks<$19 & 0.29 \\
\noalign{\smallskip}
\noalign{\smallskip} 
\hline 
\end{tabular}
\\
\end{flushleft} 
\end{table}

Data from short and long exposures were combined in order to increase the dynamical range of the 
photometry for both clusters. To do this we searched for a common set of stars in the CMD that 
minimises photometric errors and, at the same time, avoids saturation effects.

\section {The cluster DBSB\,48 in Hoffleit\,18}
\label{DBSB48}

\subsection{Cluster structure}
\label{CS}

\begin{figure}
\resizebox{\hsize}{!}{\includegraphics[angle=0]{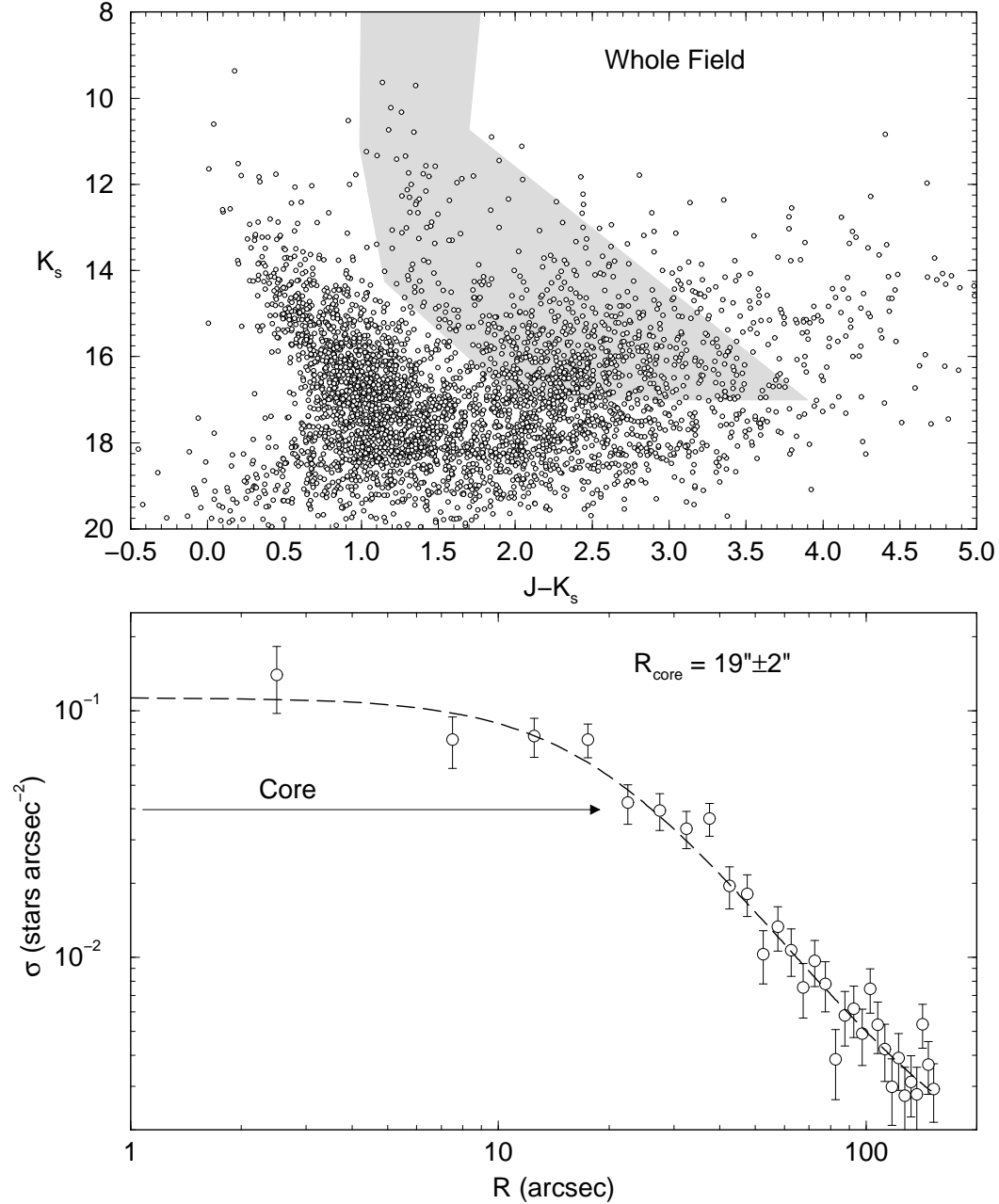}}
\caption{Top panel: Whole field $\ks\times J-\ks$ CMD in the direction of DBSB\,48. The shaded
region shows the colour-magnitude filter used to derive the cluster spatial structure. Bottom 
panel: Radial density profile of stars with the best-fit King profile superimposed. Error bars
are $1\sigma$ Poisson fluctuation.}
\label{fig3}
\end{figure}

Fig.~\ref{fig3} (top panel) shows a whole field $\ks\times J-\ks$ CMD of DBSB\,48. Different
extractions centred on DBSB\,48 show that cluster sequences are more probably contained 
in the region marked off by the colour-magnitude filter (e.g. \citealt{OldOCs};
\citealt{DenseFields}) shown in the figure (see also the central extraction in panel (a) of 
Fig.~\ref{fig4}).  The colour-magnitude filter was designed based on the decontaminated
photometry and the presence of O stars, see Sect.~\ref{FSC}. A pronounced blue main sequence 
(mostly from the disc) is seen on the left 
side of the CMD (Fig.~\ref{fig3}), and a considerable number of stars are seen to redder colours. 
Following e.g. \citet{BoBi05} and \citet{OldOCs}, we build the radial density profile (RDP) for 
stars with colours within the colour-magnitude filter. This procedure eliminates most of the stars 
with colours compatible with the field, which enhances cluster density profiles, especially in very 
crowded fields, as for the embedded open cluster NGC\,6611 \citep{BoSaBi06}. The resulting RDP is 
shown in the bottom panel of Fig.~\ref{fig3}. Despite the cluster's young age (Sect.~\ref{KEXC}), 
the RDP of DBSB\,48 follows the two-parameter \citet{King66}\footnote{$\sigma(R)=\sigma_O/(1+
(R/R_{core})^2)$, which describes the central region of star clusters.} surface density profile, with a core 
radius of $\rc=19\pm2\arcsec$. The cluster may extend at least to $r=100\arcsec$, 
since we are limited by frame dimensions. The similarity of the observed RDP with a King profile is no 
surprise for an embedded cluster, for instance, the radial distribution of stars in NGC\,6611 (${\rm age}\approx1.3\pm0.3$\,Myr) also follows this law. 

With the adopted distance of DBSB\,48 (Sect.~\ref{FP}) the absolute core radius is 
$\rc=0.46\pm0.05$\,pc, somewhat smaller than that derived for NGC\,6611 \citep{BoSaBi06}. 

The RDP can be used to select spatial extractions that minimise CMD contamination. In the subsequent 
analysis we use the regions $r\leq26\arcsec$ and $r\leq40\arcsec$ (Fig.~\ref{fig4}).

\subsection{Field-star decontamination}
\label{FSC}

To uncover the intrinsic cluster-CMD 
morphology we apply a field-star decontamination procedure. We take as comparison field the East 
border strip with dimension $7\arcsec\times300\arcsec$. This region is large enough to produce 
statistical representativity of field stars, both in magnitude and colours. This procedure was 
previously applied in the analysis of low-contrast \citep{BicBon05}, young embedded \citep{BoSaBi06}, 
and young \citep{BoBiOrBa06} OCs. 

\begin{figure}
\resizebox{\hsize}{!}{\includegraphics[angle=0]{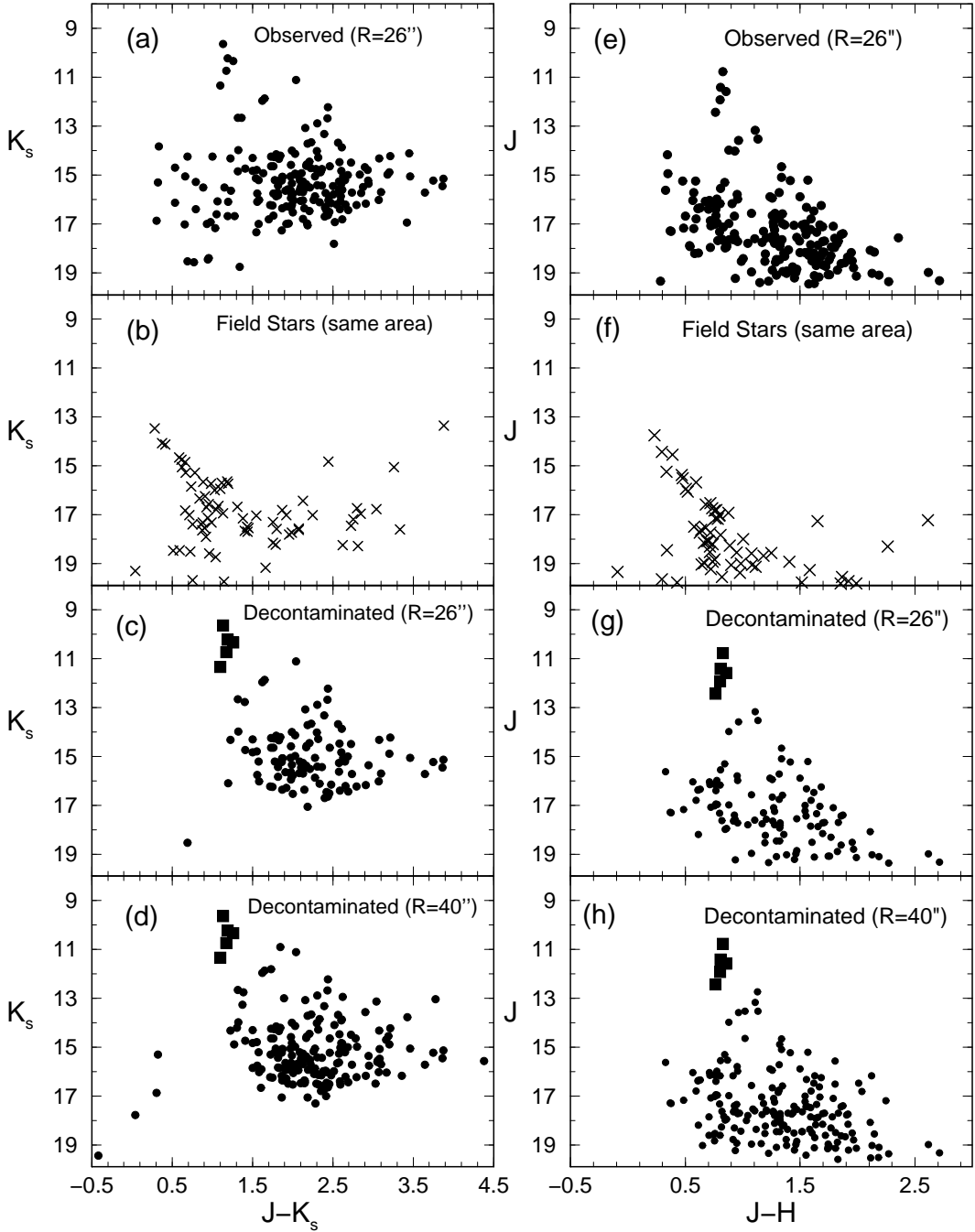}}
\caption{Panel (a): $\ks\times J-\ks$ CMD of DBSB\,48 for an extraction
of $r<90$ pixels ($r<26\arcsec$ ). Panel (b): field stars extracted from a rectangular
strip at the East border with dimension $7\arcsec\times300\arcsec$. Panel (c): decontaminated
CMD where upper-MS stars are shown as filled squares. Panel (d): decontaminated CMD of the
region $r<40\arcsec$. The corresponding $J\times J-H$ CMDs are shown in the right panels.}
\label{fig4}
\end{figure}

The algorithm works on a statistical basis that takes into account the relative densities of stars 
in a cluster region and comparison field. It {\em (i)} divides the CMD in colour/magnitude cells
of varying size, {\em (ii)} computes the expected surface density of field stars in each cell,
assuming some uniformity throughout the cluster field, and finally {\em (iii)} subtracts the expected 
number of candidate field stars from each cell. Typical sizes of the colour/magnitude cells are
$\Delta(J-\ks)=\Delta(J-H)=0.25$ and $\Delta\,J=\Delta\,\ks=0.5$, which are wide enough to allow for a
representative statistics and, at the same time, preserve the morphology of the CMD sequences. Because 
the remaining stars are in CMD cells where the stellar density presents a clear excess over the field, 
they have a high probability of being cluster members. Since field stars are taken from an external 
region of fixed area, corrections are made for the different solid angles of cluster and comparison 
field. To check consistency we apply the algorithm to the CMDs involving $\ks\times J-\ks$ and $J\times 
J-H$, separately.  As discussed in \citet{DenseFields}, what is really critical for the 
decontamination algorithm - the cell size, in particular - is the differential reddening between cluster 
and field stars. Large gradients would require large cell sizes or, in extreme cases, would preclude
application of the algorithm. However, the CMDs extracted from the cluster region and comparison 
field (Fig.~\ref{fig4}) indicate that the differential reddening in the direction of
DBSB\,48, although present, is not excessively large. In any case, some residual contamination by
very red field stars is expected to remain in the CMDs. A similar conclusion applies to Trumpler\,14 (Fig.~\ref{fig9}).

Fig.~\ref{fig4} (panel a) shows a $\ks\times J-\ks$ CMD extraction of the region $r<90$ pixels 
($r<26\arcsec$). This radius makes the CMD sequences clearer.
In fact, compared to the whole field (Fig.~\ref{fig3}, top panel) the disc MS stars 
becomes depleted with respect to that of the red sequence. This suggests that the cluster upper 
MS is present particularly around $J-\ks\approx1.0$ and $11.5<\ks<9.5$. The fainter stars appear 
to define a populated distribution around $\ks\approx14.4$ and $J-\ks\approx1.85$, however including
a considerable contamination. In addition, stars extend to very red colours up to 
$J-\ks\approx4$, which are mostly infrared excess stars (Sect.~\ref{FP}). In panel (b) we show field 
stars extracted at the frame edge with the same area as the extraction in panel (a). As expected, 
the blue sequence appears to be a field property. In panel (c) we show the decontaminated 
CMD, where the blue disc sequence vanishes, but a significant fraction of red stars remains, to be associated
to PMS stars with and without infrared emission. For comparison purposes we also show in panel 
(d) the decontaminated CMD for the extraction $r<40\arcsec$, where the bright MS sequence 
remains unchanged. This CMD suggests that the MS is still developing, 
with the faintest stars at $\ks\approx14.4$ (Sect.~3.4.1). In panels (e) to (h) we show the counterpart
$J\times J-H$ CMDs. Basically the same conclusions apply, indicating that the CMD features are
intrinsic.

\subsection{Photometric diagrams}
\label{PD}

\begin{figure}
\resizebox{\hsize}{!}{\includegraphics[angle=0]{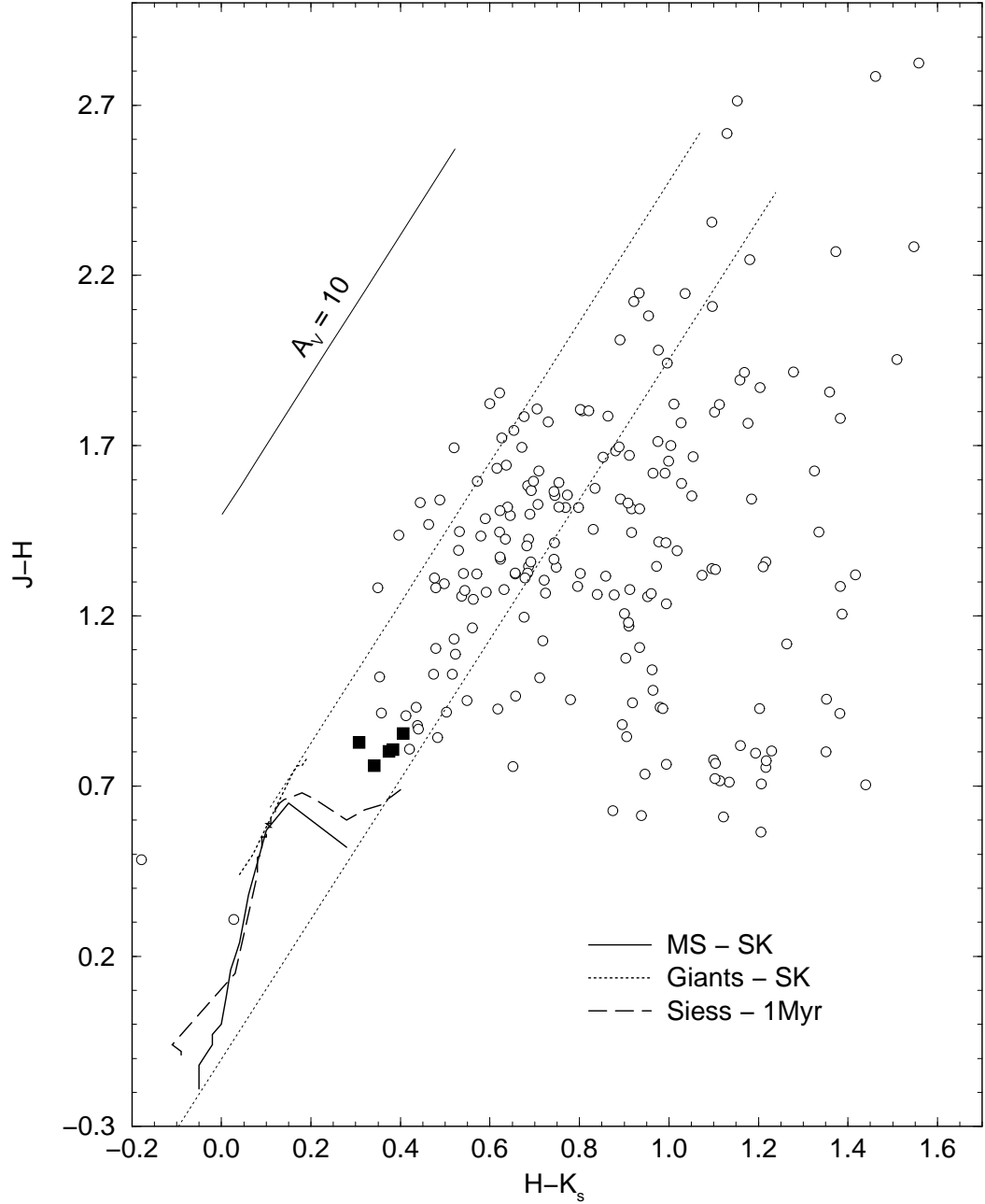}}
\caption{2-CD of DBSB\,48 using the decontaminated photometry. MS (heavy solid line) and giant 
(heavy-dotted)
tracks are from \citet{SK82}; 1\,Myr PMS track (dashed) is from \citet{SDF2000};
reddening band (thin-dotted) and $A_V=10$ reddening vector are according to \citet{KH95}.
The 5 upper-MS stars are indicated by filled squares.} 
\label{fig5}
\end{figure}

The $H-\ks\times J-H$ 2-CD of the cluster DBSB\,48 (Fig.~\ref{fig5}), was built with the 
decontaminated photometry for $r<26\arcsec$. It can be used to identify PMS stars 
with and without K-excess emission and extract information for cluster parameter 
determination. The intrinsic MS and giant colours are from \citet{SK82}, and 
the 1\,Myr  PMS track is from \citet{SDF2000}. Reddening lines are 
from \citet{KH95}. The bluer line in $H-\ks$ corresponds to reddened M5\,III stars, while the
redder one to O\,V stars.

Reddening determination for upper MS stars is based on the ZAMS's intrinsic sequence 
of early spectral types in the 2-CD (and CMD). For PMS stars it is measured taking as reference
intrinsic T Tauri colours free of K-excess, at $(H-\ks)_{\circ}=0.23$ and $(J-H)_{\circ}=0.60$
(\citealt{SKS95}; \citealt{SoBiAhCl05}). The probable PMS stars used for 
reddening calculations are those without K-excess found between the two  reddening  lines
in Fig.~\ref{fig5}. K-excess stars are those to the right of the MS reddening vector
related to O stars, and are not used for reddening determination. Instead,
they are used to constrain the cluster age (Sect~\ref{KEXC}).

Some of the stars follow the reddening vectors and are contained in the band defined
in between them, and some show K excesses. Reddening is on 
average large. For the MS stars we derive a value of $E(J-\ks)=1.40\pm0.05$, converting 
to $A_V\approx8.2\pm0.3$, and $A_{K_s}\approx0.9\pm0.03$. The uncertainty in $E(J-\ks)$
corresponds to the error of the mean. On the other hand, the spread in colour excess is
$\Delta\,E(J-\ks)\approx1$ corresponding to a range in absorption of $\Delta\,A_V=6$.

\subsection{Fundamental parameters}
\label{FP}

\begin{figure}
\resizebox{\hsize}{!}{\includegraphics{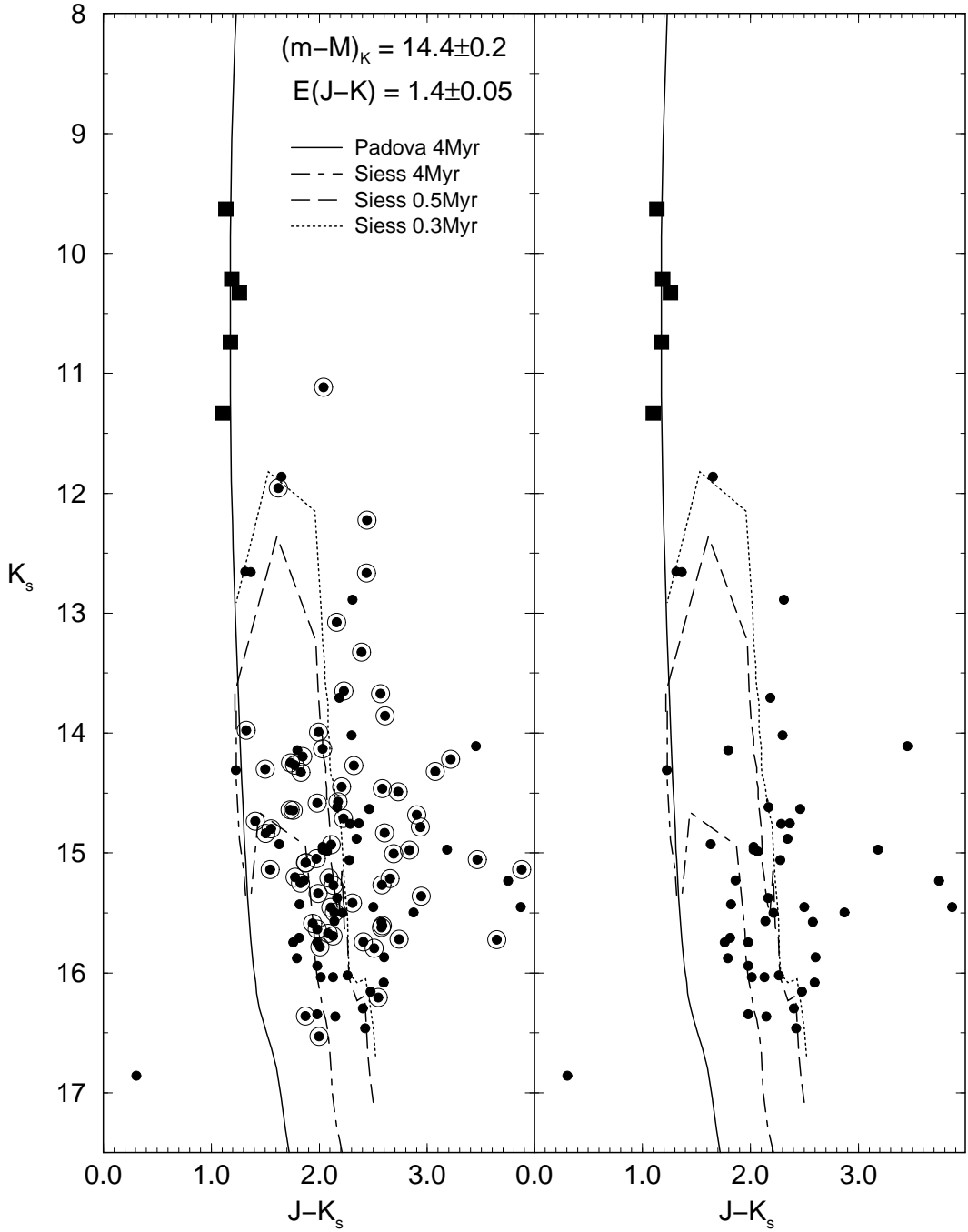}}
\caption{Left panel: decontaminated $\ks\times J-\ks$ CMD of DBSB\,48 for the
extraction $r=26\arcsec$. PMS tracks with ages 0.3, 0.5 and 4\,Myr and the 4\,Myr
solar-metallicity Padova isochrone are shown. Upper-MS stars are shown as filled squares.
K-excess stars are identified with a surrounding circle. Right panel: stars without K-excess
emission. Notice that about 90\% of the remaining stars basically follow the PMS tracks.}
\label{fig6}
\end{figure}

In Fig.~\ref{fig6} we show the decontaminated $\ks\times J-\ks$ CMD for the extraction $r=26\arcsec$. 
The lack of observational constraints precludes a precise derivation of the distance from the Sun 
using the present data. In this case, we adopt the kinematical
distance of the associated nebula Hoffleit\,18, $\ds=5.0\pm0.7$\, kpc \citep{CaHa87} to set 
the distance modulus of the isochrones. The observed and absolute distance 
moduli are $(m-M)_{K_s}=14.4\pm0.3$ and $(m-M)_{\circ}=13.48\pm0.3$. PMS tracks \citep{SDF2000} with 
the ages 0.3, 0.5 and 4\,Myr are used. Four probable turn on stars at $11.8<\ks<12.8$ basically 
coincide with the 0.3\,Myr track. In the $r=40\arcsec$  extraction 
(panel d of Fig.~\ref{fig4}) this sequence is more populated. 

We also analyse the 5 upper-MS stars in the context of distance estimate. The upper MS has a small
magnitude extent of $\Delta\,\ks = 2$, implying a very early evolutionary state for DBSB\,48. This 
extent remains unchanged on a larger extraction (panel d of Fig.~\ref{fig4}). With the adopted 
reddening and kinematic distance, the brightest star in the upper MS has $M_{K_s}=-4.8$, comparable to 
$M_{K_s}=-4.7$ of an O5V star in  \citealt{BinMer98}, and references therein. The presence of O stars 
is expected from ionisation in the nebula. Consequently, the spectral type of the other upper-MS 
stars would be around O8V, corresponding to an intrinsic $M_{K_s}=-4.0$. Alternatively, taking these 
upper-MS stars as indicators, a mean distance of $\ds=3.6\pm1.0$\,kpc is implied, consistent with 
the adopted kinematic distance, within uncertainties.
 
\begin{figure}
\resizebox{\hsize}{!}{\includegraphics{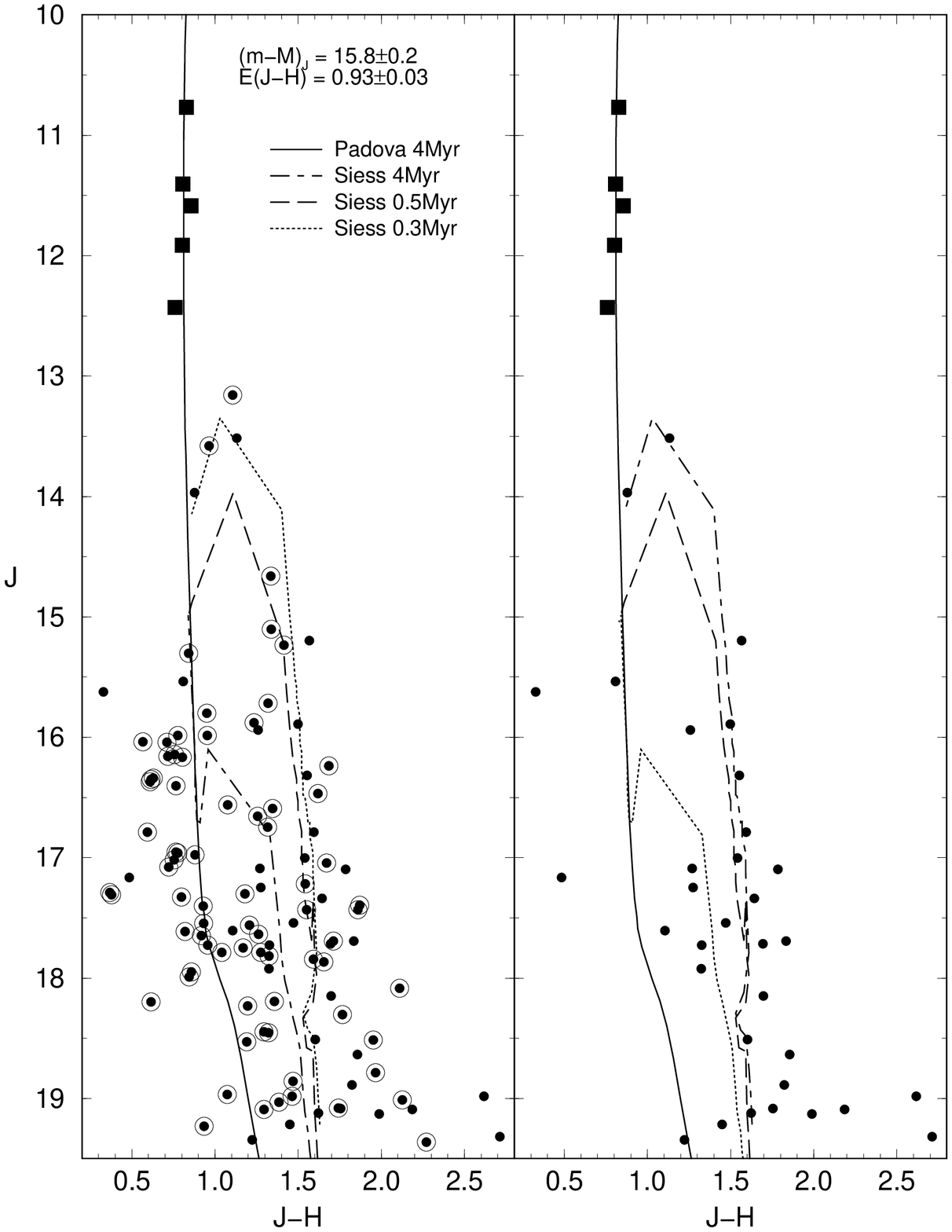}}
\caption{Same as Fig.~\ref{fig6} for the $J\times J-H$ plane (DBSB\,48).}
\label{fig7}
\end{figure}

Fig.~\ref{fig6} shows the youngest (4\,Myr) available Padova isochrone \citep{Girardi02}
set with the above parameters. MS and PMS tracks occupy consistent loci, considering that 
IR excesses affect a fraction of the PMS stars (Sect.~\ref{KEXC}).

Part of the small colour dispersion among the bright stars may be due to grain-destruction caused by 
the ionising stars. The colour range of the PMS candidates in the decontaminated photometry decreased
to about 0.5\,mag, thus within that of the PMS tracks distribution. This is not necessarily all due 
to differential reddening, since age dispersion among PMS stars is present. 

The upper-MS is consistent with any Padova isochrone younger than $\approx10$\,Myr.
However, the fact that we are dealing with an ionising nebula restricts the age to values
younger than about 5\,Myr. A more constrained age-range is provided by the PMS stars, which
suggest the presence of ages from about 0.3\,Myr to 4\,Myr. These values suggest an intrinsic 
star formation age-spread. 

To check the nature of the colour-dispersion in Fig.~\ref{fig6} (left panel) we isolate the stars 
with \ks-excess in Fig.~\ref{fig5} and identify them on Fig.~\ref{fig6} (left panel). We conclude 
that most of the $J-\ks$ colour dispersion can be accounted for by \ks-excess emission. In the right 
panel only stars without K-excess emission are shown. About 90\% of the remaining stars are found 
close to the PMS tracks, which supports a star formation age-spread, and shows as well evidence 
of differential reddening.

Fig.~\ref{fig7} adds another dimension to the analysis by the inclusion of the H band, from 
which we conclude that the results are essentially the same. The isochrones were set with values 
similar to those in Fig.~\ref{fig6}, $(m-M)_J=15.8\pm0.2$ and $E(J-H)=0.93\pm0.03$.

\subsubsection{Age from K-excess fractions}
\label{KEXC}

The fraction of stars with K-excess emission correlates inversely with cluster age (e.g. 
\citealt{HLL01b}). On theoretical grounds it is estimated that because of disc-depleting 
processes such as irradiation by the central star, viscous accretion and mass loss due to 
outflow, the median lifetime of optically thick, inner accretion discs may be as short as 
$2-3$\,Myr (\citealt{H05}). Observations indicate that significant fractions of stars 
younger than 1\,Myr 
have already lost their discs. However, they also indicate that a small fraction of stars 
$8-16$\,Myr old may retain their inner discs (e.g. \citealt{CJGB05}; \citealt{Low05}). Similar 
conclusions were reached by \citet{ACP2003} who found that $\sim30\%$ of the stars in young 
clusters lose their discs in less than 1\,Myr, while the remainder keep them for about $1-10$\,Myr.
Observational estimates of disc survival time-scales are important for planet formation theories
\citep{BZA2000}.

 The observations of DBSB\,48 (and Tr\,14) were obtained with the \ks\ filter. However,
the difference between K and \ks\ is negligible for our purposes (\citealt{DuOrBiBaZM03}; \citealt{Persson98}). In fact, K and \ks\ rarely differ more than 0.02\,mag for red  stars,
the average difference being $0.0096\pm0.017$\,mag (\citealt{Persson98}). 

For DBSB\,48 we quantify K-excess fractions by counting the number of stars with the H-\ks\ colour 
redder than the OV/late dwarfs reddening vector in the decontaminated 2-CDs (Fig.~\ref{fig5}).
We adopted the extraction
$r=26\arcsec$ for the optimal cluster/field contrast. The resulting K-fraction
is $f_K=55\pm6\%$, where the uncertainty corresponds to $1\sigma$ Poisson fluctuation. Different 
extractions produce similar results. The K-excess fraction implies an age (\citealt{BoBiOrBa06}
and references therein; see also \citealt{SoBi03} and references therein) of $1.1\pm0.5$\,Myr 
for DBSB\,48. This suggests a cluster still in the process of developing the MS (Sect.~\ref{FSC}). 



\section{Trumpler\,14}
\label{Tr14}

\subsection{Cluster structure}

\begin{figure}
\resizebox{\hsize}{!}{\includegraphics{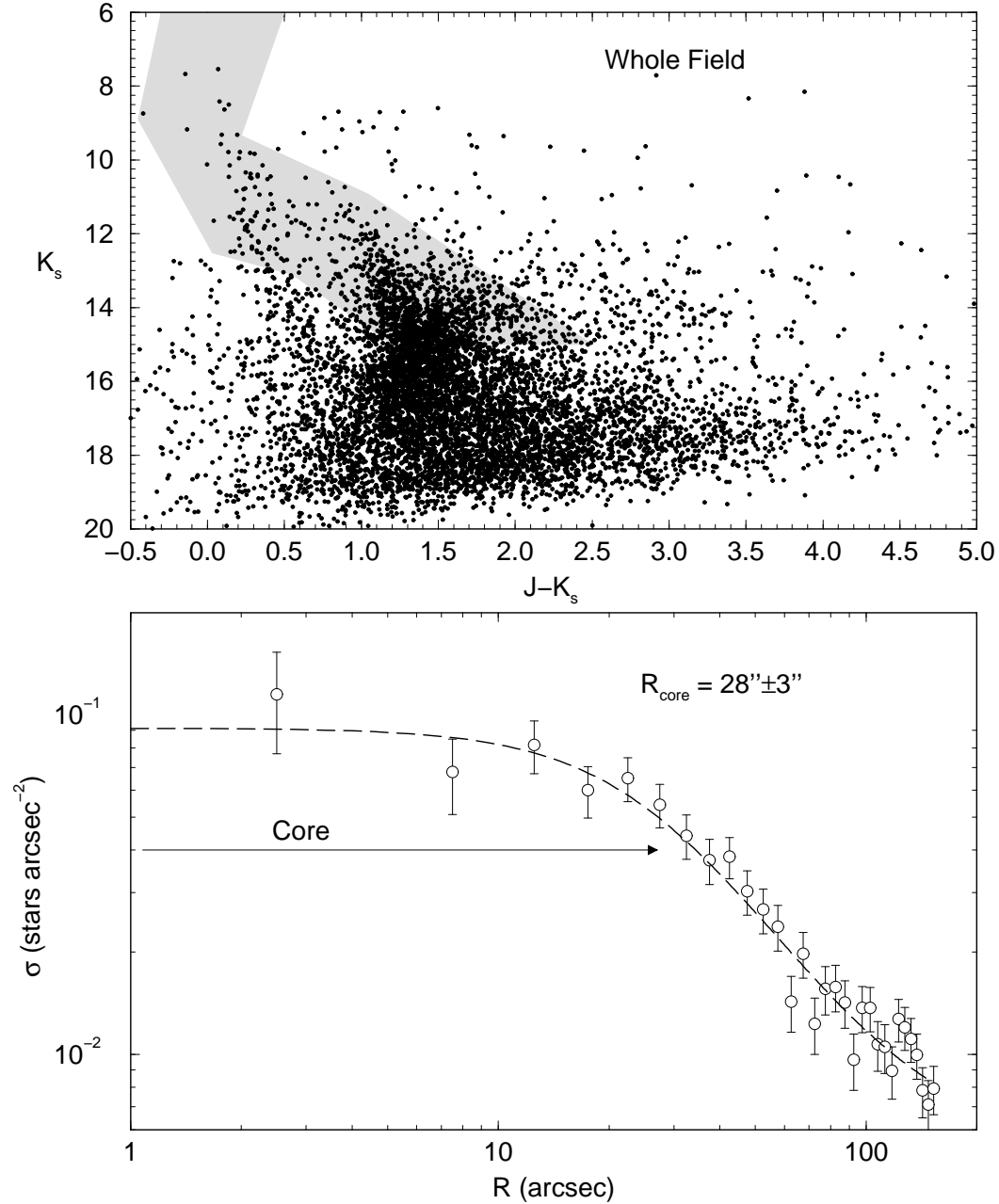}}
\caption{ Top panel: Whole field $\ks\times J-\ks$ CMD in the direction of Trumpler\,14. The 
shaded region shows the colour-magnitude filter used to derive the cluster spatial structure. 
Bottom panel: Radial density profile of stars with the best-fit King profile superimposed. Error 
bars are $1\sigma$ Poisson fluctuation. }
\label{fig8}
\end{figure}

The full-frame $\ks\times J-\ks$ CMD of Trumpler\,14 (top panel of Fig.~\ref{fig8}) shows 
the cluster's colour-magnitude filter superimposed. Contamination by field stars
smears cluster sequences, although the upper-MS can be seen 
for $\ks\leq14$. 

Similarly to DBSB\,48 (Sect.~\ref{DBSB48}) we built the radial density profile of Trumpler\,14 
(bottom panel of Fig.~\ref{fig8}) after isolating stars with the respective colour-magnitude 
filter. King's profile provides a core radius $\rc=28\pm3\arcsec$ that together with the distance
(Sect.~\ref{Tr14FP}) translates into $\rc=0.35\pm0.04$\,pc. This core is smaller than those of 
DBSB\,48 and NGC\,6611 (Sect.~\ref{CS}), again suggesting a  young age. Indeed,
cluster expansion in the first few $\sim10^7$\,yr is expected to occur following the rapid 
expulsion of the unused gas by massive winds and supernovae (e.g. \citealt{delaF2002}; \citealt{BK2002}; \citealt{GoBa06}). As a consequence, clusters expand in all scales as they 
reach for vitalisation. The cluster seems to extend beyond 100\arcsec, which implies a limiting 
radius larger than $\approx1.2$\,pc.

\subsection{Field-star decontamination}

We apply the decontamination procedure (Sect.~\ref{FSC}) to disentangle the intrinsic 
cluster-CMD morphology from field stars. To minimise the effects of crowding in the 
central parts we restricted the analysis to stars with DAOPHOT shape parameter\footnote{The 
DAOPHOT SHARP parameter is an image radius  index which is greater
than  zero if the  object is  more extended  than the expected stellar
profile  and less  than zero when   the detection appears sharper. The
expected value for a point source is zero (\citealt{Stet88}).}  $<0.6$.

\begin{figure}
\resizebox{\hsize}{!}{\includegraphics{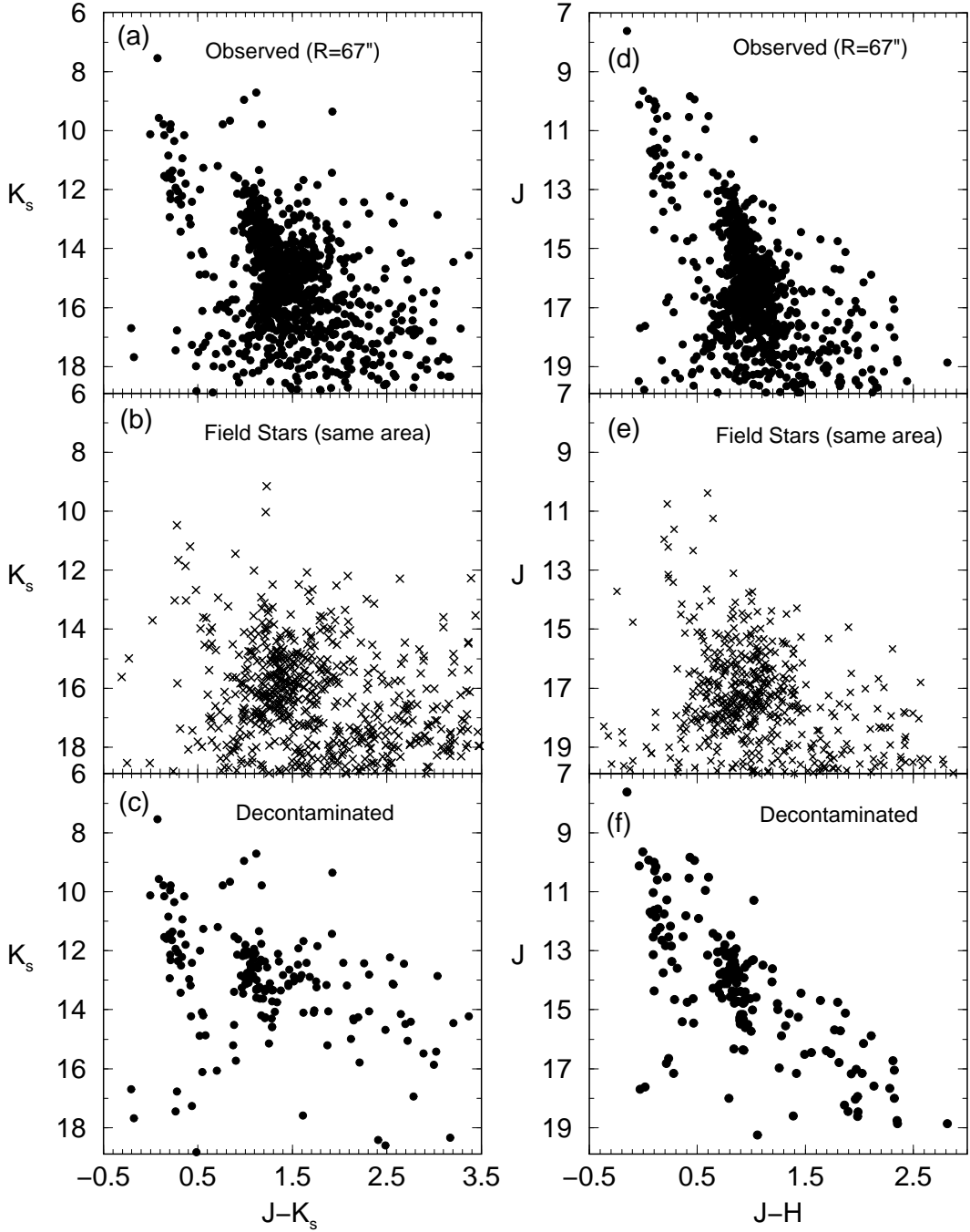}}
\caption{Same as Fig.~\ref{fig4} for Trumpler\,14 with the extraction $R=67\arcsec$. To minimise 
crowding effects the photometry was restricted to stars with DAOPHOT shape parameter $<0.6$.}
\label{fig9}
\end{figure}

Based on 
the cluster's RDP (Fig.~\ref{fig8}) we decontaminate the region $r=67\arcsec$ that corresponds to 
an optimal cluster/field contrast. We take 
as comparison field the top, bottom, left and right border frame regions with 50\arcsec\ width. The 
observed $\ks\times J-\ks$ CMD for this region (panel a of Fig.~\ref{fig9}) is compared with the 
same-area field stars (panel b). The decontaminated CMD (panel c) presents 
MS and PMS sequences.  The corresponding $J\times J-H$ CMDs are shown in the right
panels, where essentially the same features can be seen.

\subsection{Fundamental parameters}
\label{Tr14FP}

\citet{VBF96} with optical photometry obtained a mean reddening of $E(B-V)=0.56$ ($A_V=1.7$), 
a distance from the Sun $\ds= 3.1$\,kpc and an age $1.5\pm0.5$\, Myr. From the age-spread of 
PMS tracks they concluded that star formation lasted  about 5 Myr. 

\begin{figure}
\resizebox{\hsize}{!}{\includegraphics{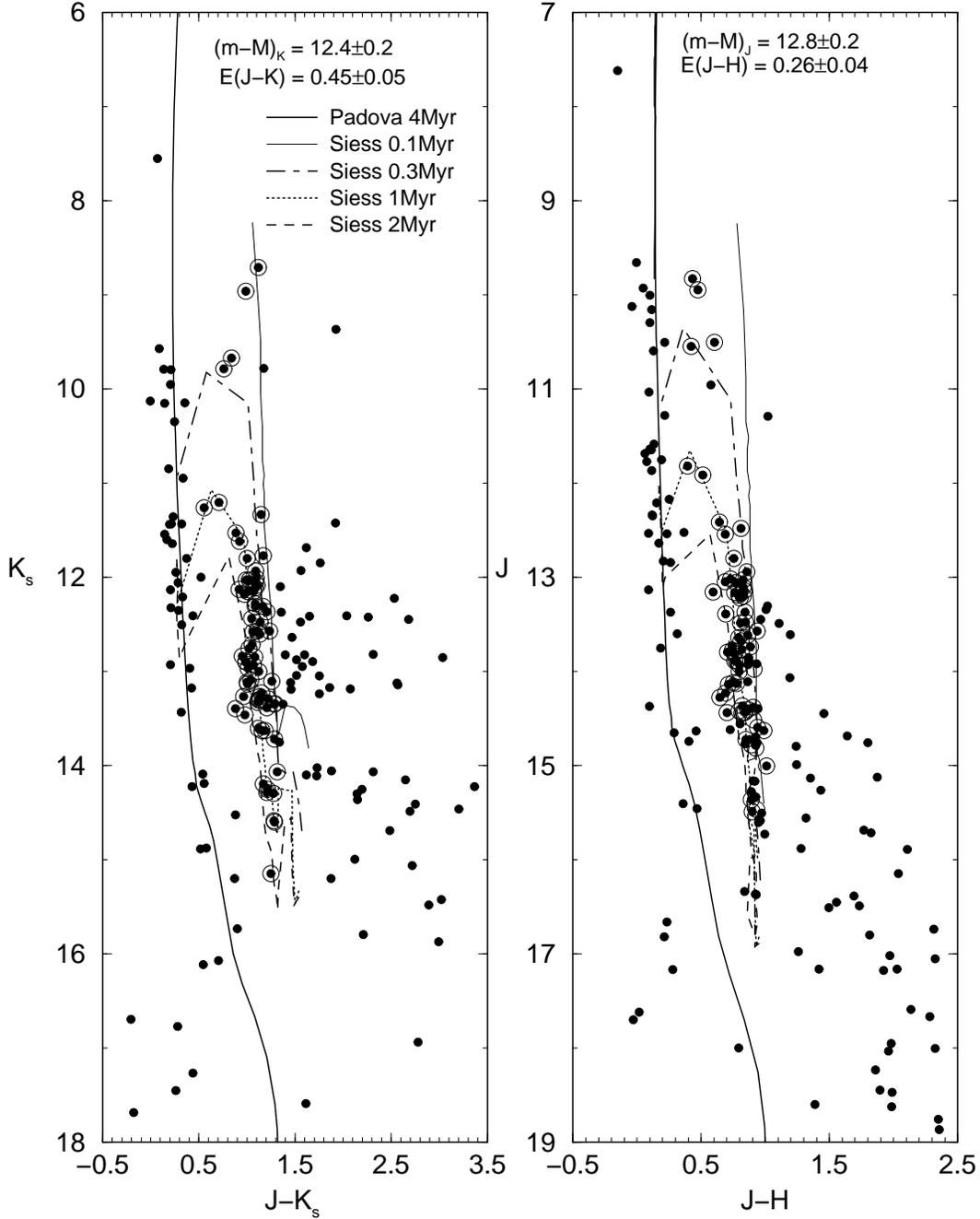}}
\caption{Decontaminated $\ks\times J-\ks$ CMD of Trumpler\,14 for the extraction $r=67\arcsec$.
PMS tracks of 0.3, 0.5 and 4\,Myr and the 4\,Myr solar-metallicity Padova isochrone are shown. 
Circles indicate candidate-PMS stars. Stars redder than the PMS tracks are probably field ones 
not subtracted.}
\label{fig10}
\end{figure}

Fig.~\ref{fig10} (left panel) shows the decontaminated $\ks\times J-\ks$ CMD of Trumpler\,14 for 
the extraction $r\leq67\arcsec$ (bottom panel of Fig.~\ref{fig9}). 
The {\em best-fit} with the 4\,Myr Padova isochrone was obtained for
$(m-M)_{K_s}=12.4\pm0.2$ and $E(J-\ks)=0.45\pm0.05$, corresponding to
$A_{K_s}=0.307\pm0.03$, which converts to $E(B-V)=0.84\pm0.09$ and $A_V=2.6\pm0.3$. With 
these values the PMS star sequences are basically reproduced with tracks in the age range 
$0.3 - 2$\,Myr (Fig.~\ref{fig10}).  These isochrones are set in the corresponding $J\times J-H$ 
CMD (right panel) after applying the equivalent values of distance modulus and reddening. Both CMDs
produce essentially the same results.  Stars redder than the PMS tracks ($J-\ks>1.7$
and $J-H>1.2$) are probably field ones not subtracted by the decontamination algorithm.

The present reddening value is somewhat larger than that derived by \citet{VBF96} in the optical. 
This appears to be due to differential reddening, since we considered in our analysis a larger number 
of PMS stars. This is also seen in the Padova isochrone fit to the MS, which is shifted to the red. An
alternative fit restricted to the upper MS provides a reddening $E(J-\ks)=0.35\pm0.04$.

The absolute distance modulus is $(m-M)_\circ=12.1\pm0.2$, and the distance from the Sun 
$\ds=2.6\pm0.3$\,kpc. The spectral type of the brightest star ($M_{K_s}=-4.9$) basically corresponds
to an O5V star, as expected from the spectral classification of the brightest stars \citep{VBF96}.

\begin{figure}
\resizebox{\hsize}{!}{\includegraphics{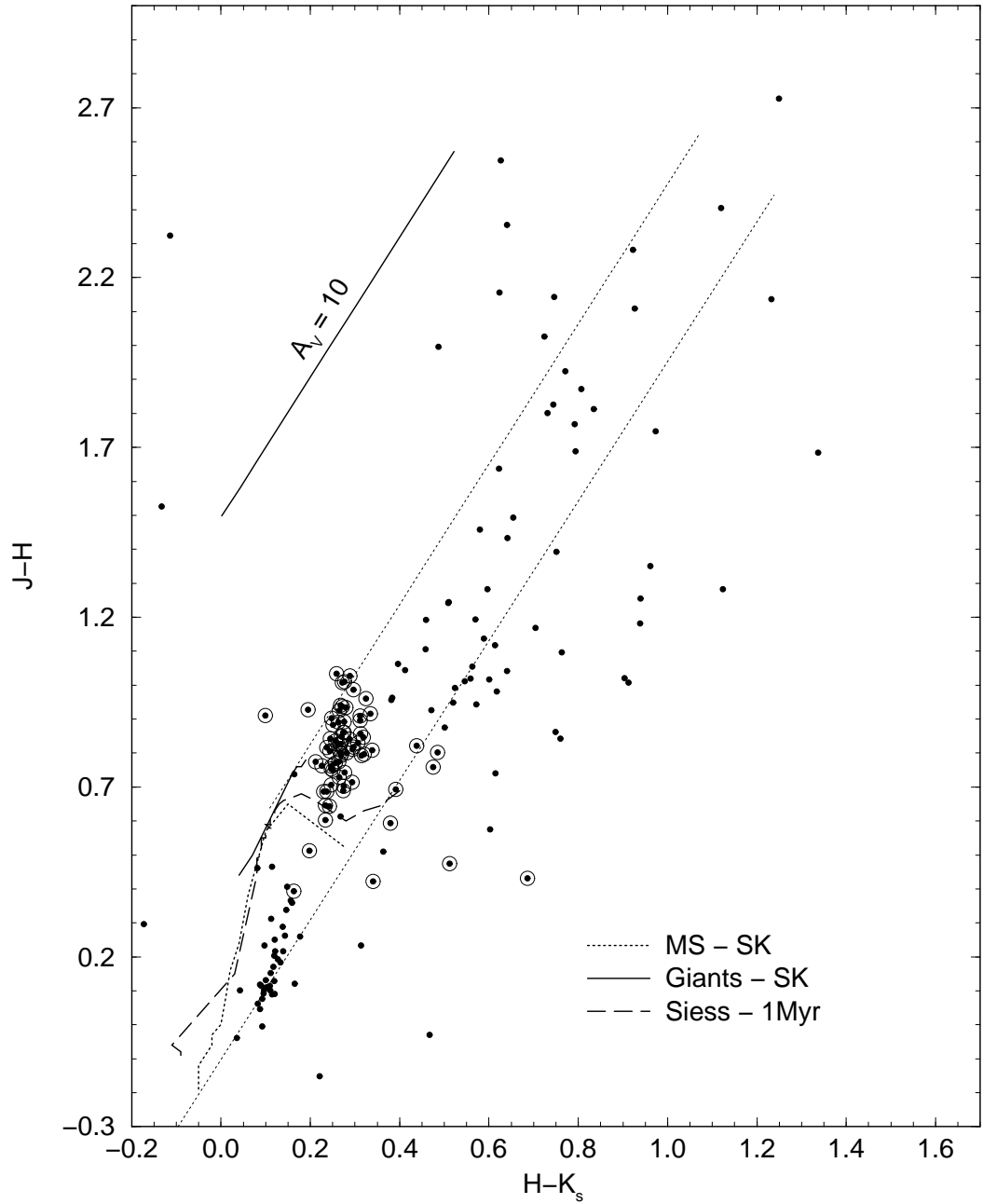}}
\caption{2-CD of Trumpler\,14 using the decontaminated photometry. MS (heavy solid line) 
and giant (heavy-dotted) tracks are from \citet{SK82}; 1\,Myr PMS track (dashed) is from 
\citet{SDF2000}; reddening band (thin-dotted) and $A_V=10$ reddening vector are according 
to \citet{KH95}. Candidate-PMS stars are indicated by circles. Field stars not subtracted
show up for $J-H>1.2$ and $H-\ks>0.6$.}
\label{fig11}
\end{figure}

Fig.~\ref{fig11} shows the 2-CD of Trumpler\,14. The region occupied by K excess stars is much 
less populated than that 
of DBSB\,48 (Fig.~\ref{fig5}). By comparing the CMD and the 2-CD, we verify that the brighter PMS 
stars populate the region $J-H\approx0.90$ and $H-\ks\approx0.28$, while the upper MS stars
are concentrated in $H-\ks\approx0.10$ and $J-H\approx0.15$.  To further check the nature
of the candidate-PMS stars we cross-identify them in Figs.~\ref{fig10} and \ref{fig11}. In
Fig.~\ref{fig10} we select the stars in the expected PMS loci. These stars are superimposed
on the 2-CD (Fig.~\ref{fig11}) where they are tightly distributed, suggesting
a common nature. Correcting their colours for $A_V=2.6\pm0.3$ (above) places them on the PMS track
in the mass range $0.7-2.2\,\ms$. In this mass range the MS coincides with the PMS, however
according to the CMDs in Fig.~\ref{fig10} we are effectively dealing with PMS stars. 
 Un-subtracted field stars show up in this diagram for $J-H>1.2$ and $H-\ks>0.6$.

For the optimal extraction $r=67\arcsec$ of Trumpler\,14 the K-fraction
is $f_K=28\pm4\%$. Similar fractions are obtained in different extractions.
Following \citealt{BoBiOrBa06} (and references therein), this K-excess fraction provides an 
age of $1.7\pm0.7$\,Myr, which is in excellent agreement with the 1.5\,Myr age 
derived in the optical \citep{VBF96}. This is consistent with the upper-MS extent
of Trumpler\,14, and supports the short MS of DBSB\,48.

\section{Discussion and Conclusions}
\label{Conclu}

 Before reaching the zero-age main sequence, stars are surrounded by optically thick material 
consisting of an infalling envelope and accretion disc that gradually disperse along the 
pre-main sequence phase. Because of disc-depleting processes such as irradiation by the central 
star, viscous accretion and mass loss due to outflow, the median lifetime of optically
thick inner accretion discs may be as short as $\rm2 - 3\,Myr$, with the final stages of disc 
accretion lasting as long as $\rm\sim10^7\,yr$ (\citealt{H05}). In addition, stars in 
young open clusters appear to form along some period of time (e.g. \citealt{Sagar95}, and 
references therein). It is in this context that the analysis of stellar density structure and 
stellar-mass distribution in young star clusters is important. 

In the present paper we studied DBSB\,48, the cluster embedded in the H\,II region Hoffleit\,18,
by means of JH\ks\ photometry, radial stellar density profiles and fraction of K-excess
emission stars. Its properties were compared to those of the young open cluster 
Trumpler\,14. Besides an important population of PMS stars of different ages,
field-star decontamination shows that the MS has developed only the upper mass range,
as expected from PMS contraction time-scales. 
The relatively short MS extent ($\Delta\,\ks=2$) of DBSB\,48 reflects its
younger age with respect to Trumpler\,14 ($\Delta\,\ks=5$). The upper MS extent
appears to be an age indicator for well populated embedded clusters.

The loci of PMS stars in DBSB\,48 are described by tracks with ages in the range 
$0.3-4$\,Myr set in the CMD with reddening $A_V=8.2\pm0.3$ and absolute distance modulus
$(m-M)_{\circ}=13.48\pm0.3$. With a fraction of K-excess emission of 
$f_K=55\pm6\%$ the age of DBSB\,48 results $1.1\pm0.5$\,Myr. Its radial density 
profile is well represented by a King profile with a core radius $\rc=0.46\pm0.05$\,pc.  

For Trumpler\,14 we derived  $\rc=0.35\pm0.04$\,pc. PMS stars are described by
$0.1-2$\,Myr tracks, consistent with the $1.7\pm0.7$\,Myr age implied by
$f_K=28\pm4\%$. 

The PMS age spread suggests that star formation in both clusters did not occur as an 
instantaneous event, instead it lasted a time equivalent to about the cluster ages.

The core radii of DBSB\,48 and Trumpler\,14  are similar to that of the embedded
open cluster NGC\,6611 and significantly smaller than those of classical open
clusters (e.g. \citealt{BoBi05}; \citealt{OldOCs}). This suggests that core formation 
is a process partly primordial, probably associated with parent molecular 
cloud fragmentation, and partly related to subsequent internal dynamical evolution. 

\section*{acknowledgements}
We thank the referee for helpful suggestions.
We acknowledge partial financial support from the Brazilian agencies Fapesp and CNPq,
and the Italian Ministero dell'Universit\`a e della Ricerca Scientifica e Tecnologica 
(MURST) under the program on 'Fasi iniziali di Evoluzione dell'Alone e del Bulge Galattico'.


\end{document}